\documentclass[10pt,conference]{IEEEtran}
\IEEEoverridecommandlockouts

\usepackage{amsmath,amsfonts}
\usepackage{array}
\usepackage[caption=false,font=scriptsize,labelfont=rm,textfont=rm]{subfig}
\usepackage{textcomp}
\usepackage{stfloats}
\usepackage{url}
\usepackage{verbatim}
\usepackage{graphicx}
\usepackage{cite}
\usepackage{gensymb}
\usepackage{CJKutf8}
\usepackage{amsthm,amsmath,amssymb} 
\usepackage{algpseudocode}   
\usepackage{mathrsfs}
\usepackage{multicol}
\usepackage{svg}
\usepackage{graphicx}
\usepackage[ruled,linesnumbered]{algorithm2e}
\usepackage{algpseudocode}
\def\BibTeX{{\rm B\kern-.05em{\sc i\kern-.025em b}\kern-.08em
    T\kern-.1667em\lower.7ex\hbox{E}\kern-.125emX}}
    
\begin{document}

\title{Near-Field Variable-Width Beam Coverage and Codebook Design for XL-RIS
\thanks{This paper was partially funded by the National Key R \& D
Program of China (2020YFB1806602), BUPT-China Unicom Joint Innovation Center and Fundamental Research Funds for the Central Universities (2242022k60006). \textit{(Corresponding author: Qiang Wang, Qiuyan Liu.)}}
}

 \author{\IEEEauthorblockN{Yida Zhang\IEEEauthorrefmark{1}, Qiuyan Liu\IEEEauthorrefmark{2}, Qiang Wang\IEEEauthorrefmark{1}, Hongtao Luo\IEEEauthorrefmark{1}, Yuqi Xia\IEEEauthorrefmark{1},}
 
 \IEEEauthorblockA{\IEEEauthorrefmark{1}
National Engineering Research Center for Mobile Network Technologies,\\
Beijing University of Posts and Telecommunications, Beijing 100876, China, \\
\IEEEauthorrefmark{2} China United Network Communications Corporation Research Institute, Beijing 100037, China\\
Email: \{zhangyida02, wangq, mashirokaze1971, xiayuqi\}@bupt.edu.cn, liuqy95@chinaunicom.cn}
}

\maketitle

\begin{abstract}
To mitigate the issue of limited base station coverage caused by severe high-frequency electromagnetic wave attenuation, Extremely Large Reconfigurable Intelligent Surface (XL-RIS) has garnered significant attention due to its high beam gain. However, XL-RIS exhibits a narrower beam width compared to traditional RIS, which increases the complexity of beam alignment and broadcast. To address this problem, we propose a variable-width beam generation algorithm under the near-field assumption and apply it to the near-field codebook design for XL-RIS. Our algorithm can achieve beam coverage for arbitrarily shaped codeword regions and generate a joint codebook for the multi-XL-RIS system. The simulation results demonstrate that our proposed scheme enables user equipment (UE) to achieve higher spectral efficiency and lower communication outage probability within the codeword region compared to existing works. Furthermore, our scheme exhibits better robustness to codeword region location and area variations.
\end{abstract}


\section{Introduction}
To utilize broader spectrum resources, 6th Generation Wireless Systems (6G) are shifting towards higher frequency bands, such as millimeter wave and terahertz bands. Unfortunately, the pronounced attenuation of high-frequency electromagnetic waves results in limited base station (BS) coverage areas~\cite{8387211}. Reconfigurable Intelligent Surface (RIS) is a novel relay device composed of subwavelength units with tunable electromagnetic response capabilities. By precisely controlling its electromagnetic (EM) elements, RIS can alter the amplitude, phase, and polarization characteristics of electromagnetic waves, thereby enabling secondary active beamforming~\cite{11091527},~\cite{2018Space}. Extremely Large RIS (XL-RIS) is a further development of RIS technology. It typically consists of thousands to tens of thousands of EM elements. Therefore, both their array gain and near-field range are significantly greater than those of traditional passive RIS~\cite{10663714}.

Codebooks are predefined sets of beams that antenna arrays use to efficiently generate beams in different directions. In 5G NR, 3GPP TS 38.211 specifies the \textit{Type I} wide beam codebook intended for large coverage and the \textit{Type II} narrow beam codebook intended for precise alignment. This dual design alleviates the degradation of alignment accuracy and the Synchronization Signal Block (SSB) broadcast difficulties caused by the excessively narrow beam width~\cite{3gpp.TS38.211}. However, the wide beam codeword of \textit{Type I} is designed based on the far field model, which inevitably leads to degradation in beamforming performance due to mismatch with the near field model~\cite{9903389}. Given this, XL-RIS critically requires a new near-field variable-width beam codebook design~\cite{IMT-2030_6G_Metasurface_Report_2024}.

\begin{figure}[!t]
\centering
\setlength{\abovecaptionskip}{-0.17cm}
\includegraphics[width=3in]{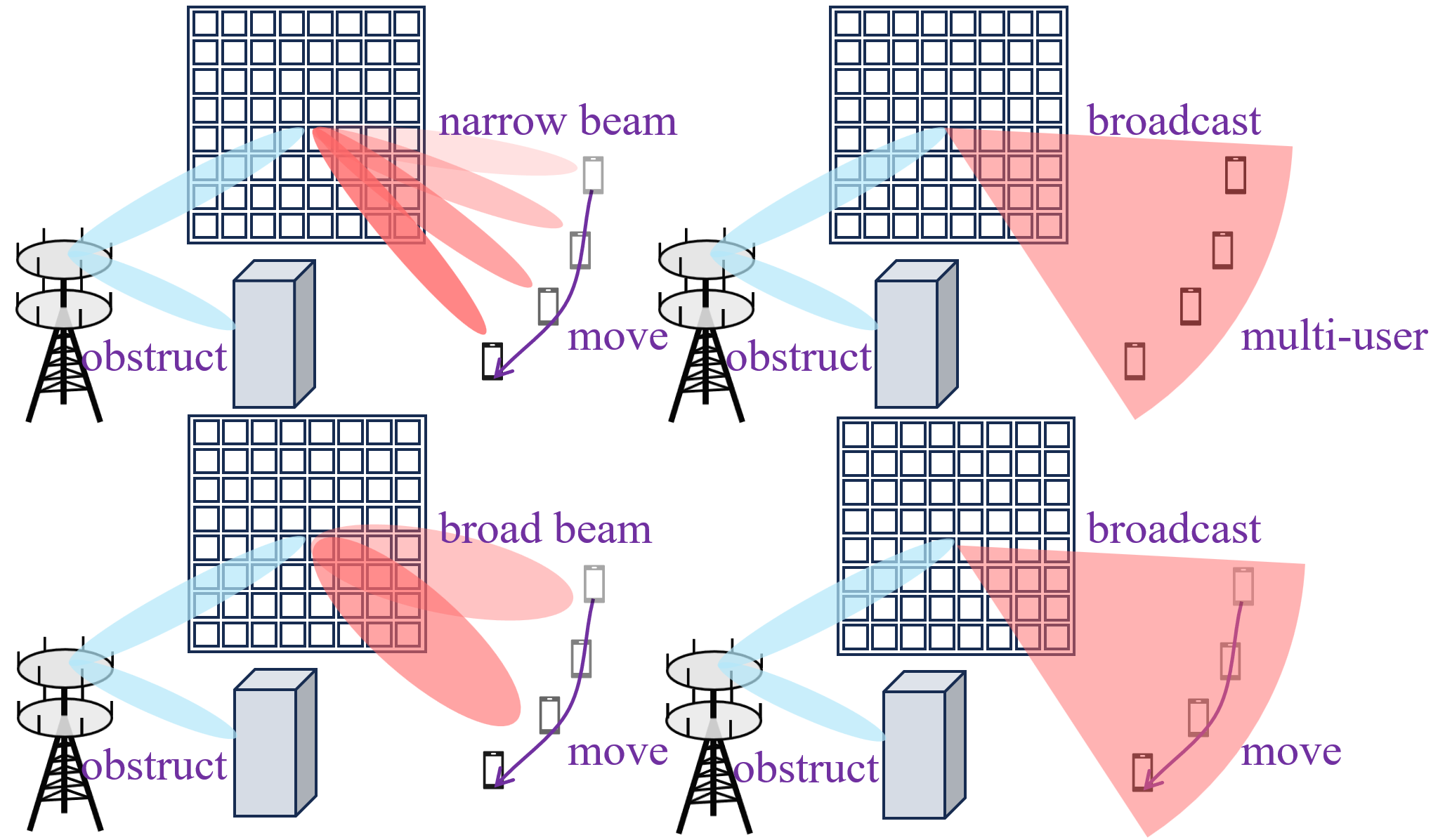}
\caption{Illustration of the application of the XL-RIS variable-width beam.}
\label{Fig1}
\vspace{-0.1cm}
\end{figure}

In recent years, some studies on near-field variable-width beamforming for RIS have been published\cite{9206044},\cite{10570947},\cite{9978148},\cite{9673796},\cite{9827873},\cite{10437101}. Reference \cite{9206044} proposed utilizing phase compensation to virtually fabricate a mirror source, thereby realizing wide beam broadcasting in various directions within the near-field. References \cite{10570947} and \cite{9978148} combined the Fresnel principle with the DFT codebook to construct a low-complexity near-field ring-type codebook. Their approach can broaden the RIS beam and flexibly switch between wide and narrow beams. References \cite{9673796} and \cite{9827873} subdivided the codeword region and utilized the mapping relationship between EM units and subregions to design a rectangular near-field variable-width beam coverage codebook. Furthermore, they investigated the application of this codebook in mobile user equipment (UE) beam management and hierarchical beam management. Based on \cite{9673796} and \cite{9827873}, Reference \cite{10437101} built an equivalent spherical model and employed the spherical mapping method to design a sector near-field variable-width beam coverage codebook. However, existing near-field variable-width beam coverage designs rely on predefined mapping relationships or design criteria, resulting in their coverage regions being limited to only a single rectangle or a single sector. Therefore, they can't achieve beam coverage for arbitrary target regions\cite{IMT-2030_6G_Metasurface_Report_2024}.

To address the aforementioned limitation, we propose a variable-width beam generation algorithm under the near-field assumption and apply it to the near-field codebook design for XL-RIS. The main contributions of this paper are as follows:
\begin{enumerate}[]
    \item We propose an algorithm for XL-RIS that provides beam coverage for the specified target area within the near-field. The algorithm is not only flexible for application in target areas of arbitrary shapes but also simultaneously covers multiple target areas. Furthermore, it supports the joint control of multiple XL-RISs to achieve cooperative beam coverage.
    \item Based on the proposed algorithm, we introduce a codebook generation framework for hierarchical beam coverage. For any given pre-defined BS codebook, this framework can generate complementary XL-RIS codewords in parallel, ultimately forming an XL-RIS codebook.
    \item Simulation results demonstrate that our codebook outperforms other variable-width beam codebooks in terms of overall beam coverage performance. Even when the angle between the codeword region and the XL-RIS normal direction is large, our scheme maintains effective beam coverage while avoiding the distortion phenomenon observed in existing designs.
\end{enumerate}

Notations: Scalars are denoted using regular typeface, vectors and matrices are represented in boldface. The sets of complex and real numbers are denoted by $\mathbb{C}$ and $\mathbb{R}$, respectively. $\mathbb{E} \left[ \cdot \right] $ denotes the expectation of a value. The transpose and trace operators are denoted by $\left( \cdot \right) ^{\mathrm{T}}$ and $\mathrm{tr}\left( \cdot \right)$. $\mathbf{X}_{:,j}$ denotes the $j$-th column of matrix $\mathbf{X}$. The entry in the n-th row and the m-th column of a matrix $\mathrm{X}$ is denoted by $\left[ \mathbf{X} \right] _{n,m}$. A complex number's modulus and principal argument are denoted by $\left| \cdot \right|$ and $\angle \left( \cdot \right) $ respectively. $Y\gg X$ denotes that $Y$ is much greater than $X$. 

\section{System Model}
\subsection{Scenarios, Assumptions, and Communication Model}
\begin{figure}[!t]
\centering
\setlength{\abovecaptionskip}{-0.13cm}
\includegraphics[width=3.5in]{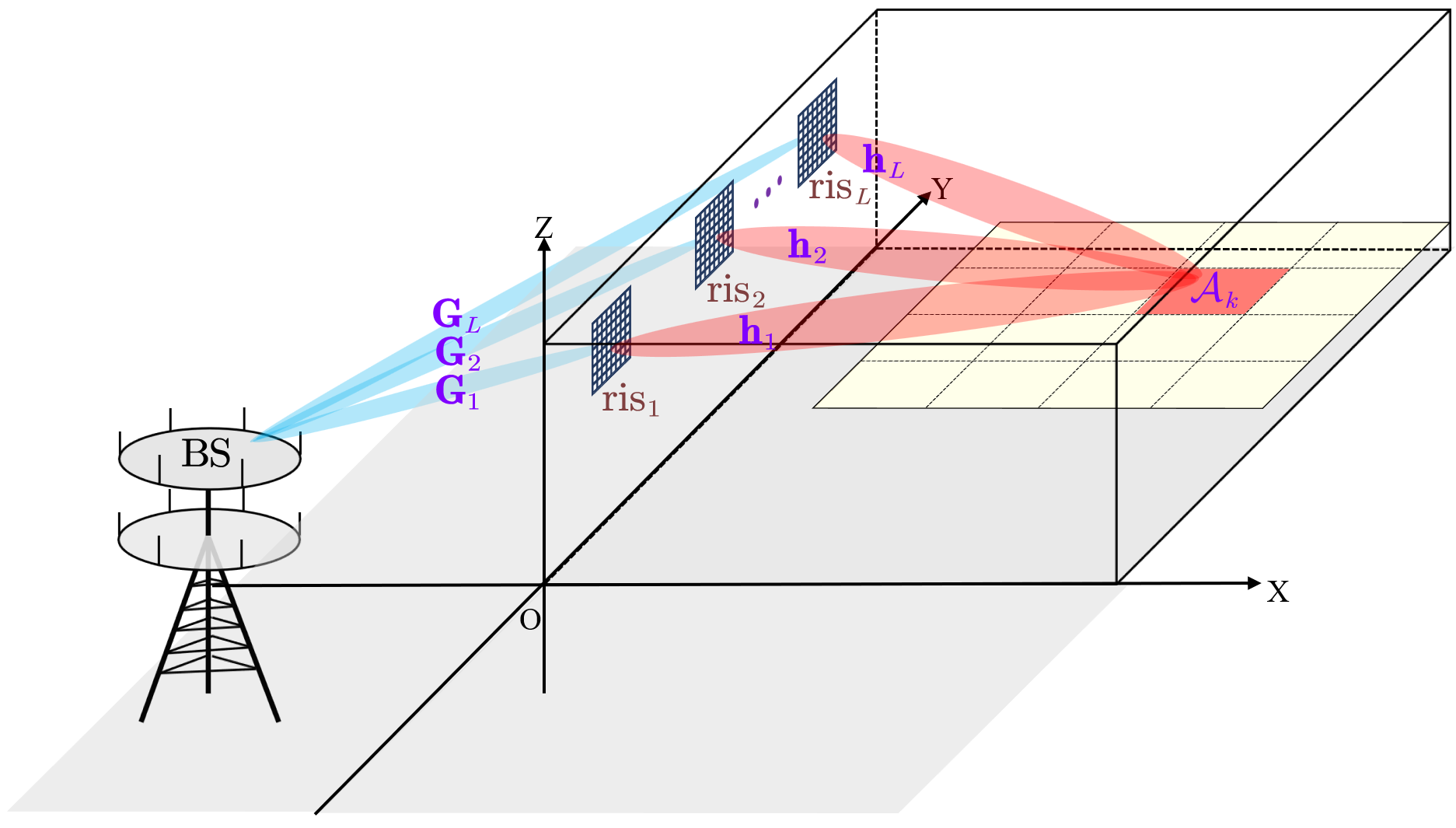}
\caption{Scenario diagram of XL-RISs-assisted communication.}
\vspace{-0.1cm}
\label{Fig2}
\end{figure}

Fig. \ref{Fig2} shows an indoor downlink scenario, where multiple transmissive XL-RISs are rationally deployed on windows to enhance indoor wireless signal coverage. The indoor UE is located in the near fields of the XL-RISs. Suppose that there are $L
$ XL-RISs located in the YOZ plane. The set of XL-RISs is denoted as $\mathcal{R}=\left\{\mathrm{ris}_1,\mathrm{ris}_2,...,\mathrm{ris}_L\right\} $. Each $\mathrm{ris}_l \in \mathcal{R}$ has $N_l$ EM units, and the total number of EM units is $N$. The transmission coefficient matrix of $\mathrm{ris}_l$ is given by
\begin{align}
\label{eq1}
&\mathbf{\Phi }_l=\mathrm{diag}\left(e^{-j\theta _1},\cdots ,e^{-j\theta _{N_l}} \right) ,\tag{1}\\
&\theta _n\in \mathcal{Q},1\leqslant n\leqslant N_l,\tag{2}\\
&\mathcal{Q}=\begin{cases}
	\left\{ \small{0,\frac{2\pi}{2^q},\cdots ,\frac{\left( 2^q-1 \right) \times 2\pi}{2^q}} \right\},if \ \mathrm{discrete},\\
	\left[ 0,2\pi \right) ,if\ \mathrm{continuous},\\
\end{cases}\tag{3}
\label{eq3}
\end{align}
where $q$ is the quantized bits of the phase. $\theta_n$ is the phase shift coefficient of the $n$-th EM element of $\mathrm{ris}_l$.

In our system model, BS is equipped with $M$ antennas. $\mathbf{C}^{\mathrm{BS}} \in \mathbb{C} ^{M\times N^{\mathrm{BS}}}$ is the codebook used by the BS which contain $N^{\mathrm{BS}}$ codewords. $\mathbf{G}_l\in \mathbb{C} ^{N_l\times M}$ and $\mathbf{h}_l\in \mathbb{C} ^{1\times N_l}$ are the channel coefficient matrices of BS to $\mathrm{ris}_l$ and $\mathrm{ris}_l$ to UE, respectively. Due to obstruction of the building, the signal can only reach the indoor UE through the XL-RISs. Thus the signal received by the UE at position $\boldsymbol{u}=\left[ x^{\mathrm{UE}},y^{\mathrm{UE}},z^{\mathrm{UE}} \right] ^{\mathrm{T}}$ is
\begin{align*}
\label{eq6}
y_{\boldsymbol{u}}=\sum_{l=1}^L{\mathbf{h}_{l}\mathbf{\Phi }_l\mathbf{G}_{l}\mathbf{C}_{:,j}^{\mathrm{BS}} \sqrt{P^{\mathrm{tx}}}s}+n_{\boldsymbol{u}},\tag{4}
\end{align*}
where $\mathbf{C}_{:,j}^{\mathrm{BS}}$,$\ \left\| \mathbf{C}_{:,j}^{\mathrm{BS}} \right\| ^2=1$ is the $j$-th codeword of the BS codebook. $P^{\mathrm{tx}}$ is the total transmit power of the BS. $s$ is the information symbol that satisfies $E\left\{ \left| s \right|^2 \right\} =1$. $n_{\boldsymbol{u}}\sim \mathcal{C} \mathcal{N} \left( 0,\sigma ^2 \right) $ is additive white Gaussian noise. We define the equivalent channel coefficient matrices $\mathbf{h}$, $\mathbf{\Phi}$, and $\mathbf{G}$ as
\begin{align*}
\label{eq5}
\left\{ \begin{array}{l}
	\mathbf{h}\overset{def}{=}\left[ \mathbf{h}_1,\mathbf{h}_2,\cdots ,\mathbf{h}_L \right] \,\in \mathbb{C} ^{1\times N},\\
	\mathbf{\Phi }\overset{def}{=}\mathrm{diag}\left( \mathbf{\Phi }_1,\mathbf{\Phi }_2,\cdots ,\mathbf{\Phi }_L \right) \,\,\in \mathbb{C} ^{N\times N},\\
	\mathbf{G}\overset{def}{=}\left[ \mathbf{G}_{1}^{\mathrm{T}},\mathbf{G}_{2}^{\mathrm{T}},\cdots ,\mathbf{G}_{L}^{\mathrm{T}} \right] ^{\mathrm{T}}\,\,\in \mathbb{C} ^{N\times M}.\\
\end{array} \right. \tag{5}
\end{align*}
Then, the useful signal power $P_{\boldsymbol{u}}^{\mathrm{rx}}$ received by the UE is
\begin{align*}
\label{eq6}
P_{\boldsymbol{u}}^{\mathrm{rx}}=P^{\mathrm{tx}}\left| \mathbf{h}\mathbf{\Phi G}\mathbf{C}_{:,j}^{\mathrm{BS}} \right|^2.\tag{6}
\end{align*}

\subsection{Channel Model}
\begin{figure}[!t]
\centering
\setlength{\abovecaptionskip}{-0.15cm}
\includegraphics[width=3.5in]{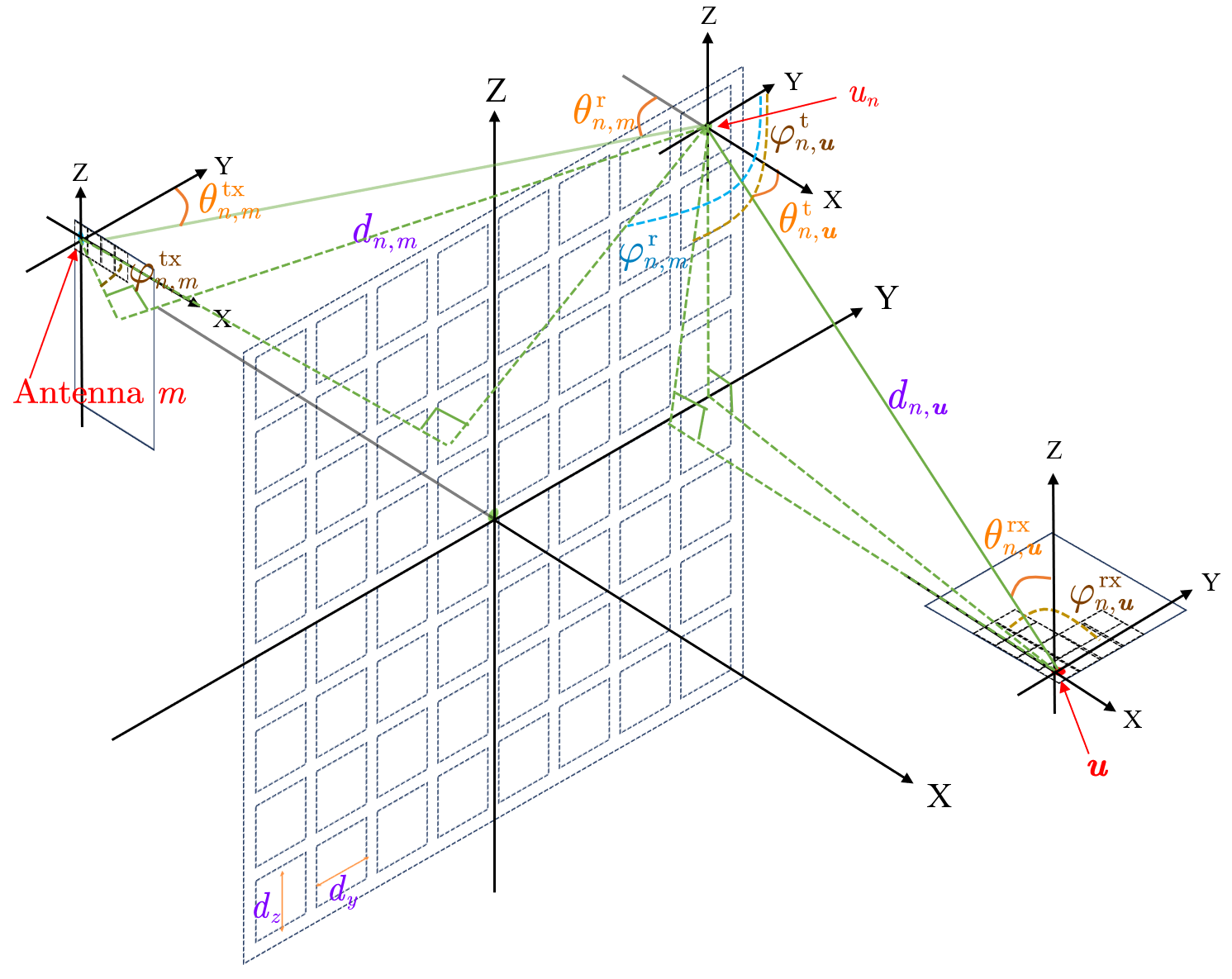}
\caption{Schematic diagram of EM unit analysis in XL-RIS.}
\label{Fig3}
\vspace{-0.05cm}
\end{figure}

We first sort all EM units according to their distance from the BS center in ascending order. Then, as shown in Fig. ~\ref{Fig3}, we perform electromagnetic analysis on the $n$-th EM unit $u_n$,  $1\leqslant n\leqslant N$. The power received by the UE is transmitted via 
$u_n$ can be expressed as~\cite{9828501},~\cite{9206044},~\cite{9541182}
\begin{align*}
\label{eq7}
P_{n,\boldsymbol{u}}^{\mathrm{rx}}=P_{n}^{\mathrm{t}}\frac{G^{\mathrm{RIS}}F_{n,\boldsymbol{u}}^{\mathrm{t}}F_{n,\boldsymbol{u}}^{\mathrm{rx}}}{4\pi \left( d_{n,\boldsymbol{u}} \right) ^2}A^{\mathrm{rx}},\tag{7}
\end{align*}
where $P_{n}^{\mathrm{t}}$ is the transmitted power of $u_n$. $d_{n,\boldsymbol{u}}$ is the distance between $u_{n}$ and UE. $G^{\mathrm{RIS}}$ is antenna radiation gain of the EM unit. $F_{n,\boldsymbol{u}}^{\mathrm{t}}$ is the normalized power radiation coefficient of $u_n$ for transmission to the UE. $F_{n,\boldsymbol{u}}^{\mathrm{rx}}$ is the normalized power radiation coefficient of the UE to receive the signal from $u_n$. $A^{\mathrm{rx}}$ is the effective aperture of UE receiving antenna. Then, the equivalent channel coefficient matrix $\mathbf{h}$ can be modeled using line of sight (LOS) as~\cite{9828501},~\cite{9206044},~\cite{9978148},~\cite{9673796},~\cite{10437101}
\begin{align*}
\label{eq8}
\left[ \mathbf{h} \right] _n=\frac{\lambda 
\sqrt{G^{\mathrm{RIS}}G^{\mathrm{rx}}F_{n,\boldsymbol{u}}^{\mathrm{t}}F_{n,\boldsymbol{u}}^{\mathrm{rx}}}}{4\pi d_{n,\boldsymbol{u}}}e^{-j{\frac{2\pi}{\lambda}}d_{n,\boldsymbol{u}}},\tag{8}
\end{align*}
where $\lambda$ is the wavelength. $G^{\mathrm{rx}}=\frac{4\pi A^{\mathrm{rx}}}{\lambda ^2}$ is the gain of the receiver antenna of UE. 

Similarly, the equivalent channel coefficient matrix $\mathbf{G}$ can also be modeled as
\begin{align*}
\label{eq9}
\left[ \mathbf{G} \right] _{n,m}=\frac{\lambda \sqrt{G^{\mathrm{RIS}}G^{\mathrm{tx}}F_{n,m}^{\mathrm{r}}F_{n,m}^{\mathrm{tx}}}}{4\pi d_{n,m}}e^{-j\small{\frac{2\pi}{\lambda}}d_{n,m}},\tag{9}
\end{align*}
where $G^{\mathrm{tx}}$ is the BS antenna gain. $F_{n,m}^{\mathrm{tx}}$ is the normalized power radiation coefficient of the $m$-th, $1 \le m \le M$ transmit antenna directed towards $u_n$. $F_{n,m}^{\mathrm{r}}$ is the normalized power radiation coefficient of $u_n$ to receive the signal from the $m$-th BS antenna. $d_{n,m}$ is the distance from the $m$-th antenna to $u_n$.

\subsection{Problem Posing}

\begin{figure}[!t]
\centering
\setlength{\abovecaptionskip}{-0.1cm}
\includegraphics[width=3.5in]{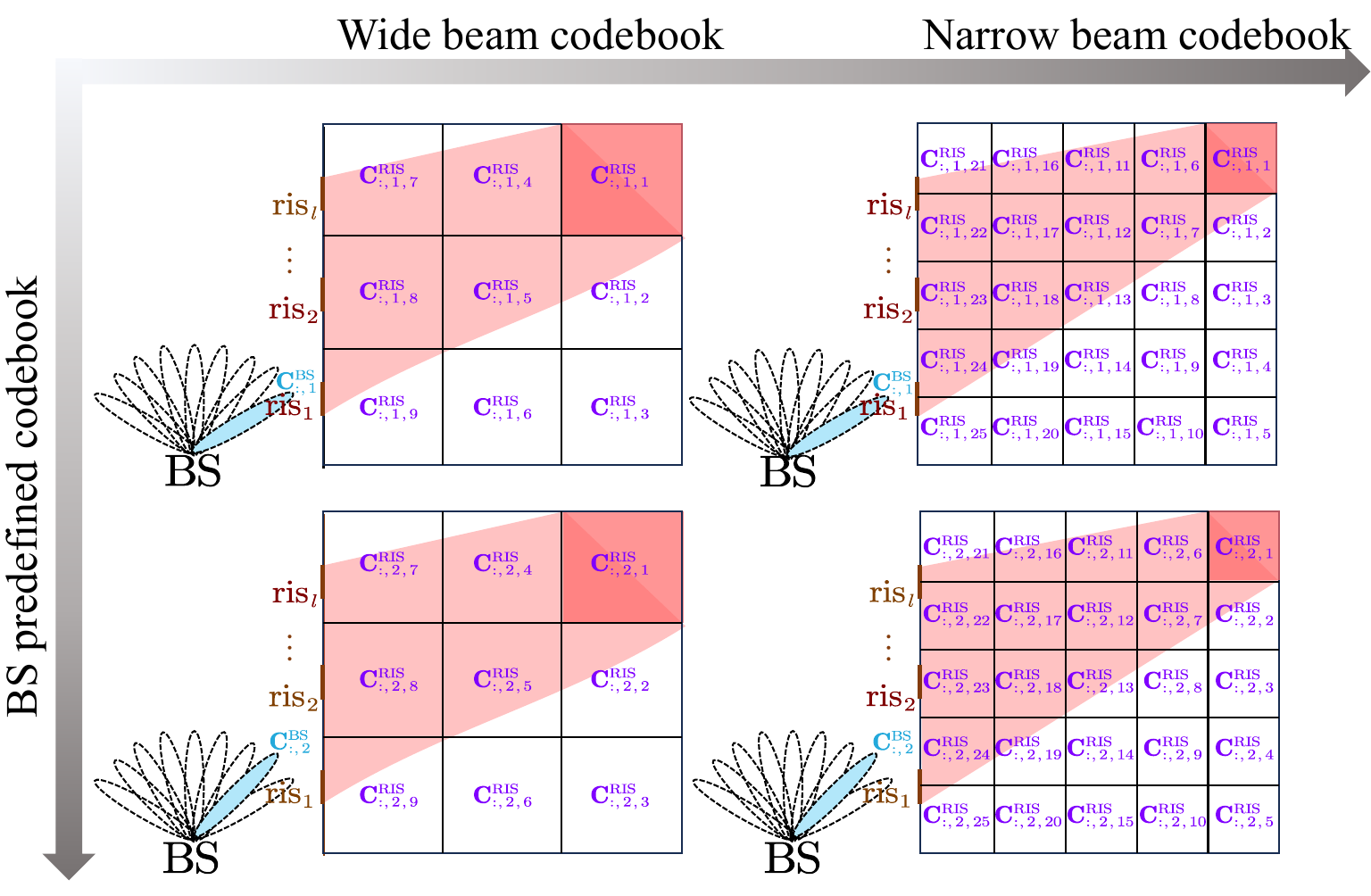}
\caption{Framework for layered beam coverage codebook.}
\label{Fig4}
\vspace{-0.1cm}
\end{figure}

We denote our designed codebook as $\mathbf{C}^{\mathrm{RIS}}\in \mathbb{C} ^{N\times N^{\mathrm{BS}}\times N^{\mathrm{RIS}}}$, which contains $N^{\mathrm{BS}}\times N^{\mathrm{RIS}}$ codewords. As shown in Fig. \ref{Fig4}, when the BS uses codeword $\mathbf{C}_{:,j}^{\mathrm{BS}}$, $1\leqslant j\leqslant N^{\mathrm{BS}}$ for beamforming, the XL-RISs can use codeword $\mathbf{C}_{:,j,k}^{\mathrm{RIS}}$, $1\leqslant k\leqslant N^{\mathrm{RIS}}$ to achieve coverage of the target area $\mathcal{A}_k$. We define $\eta _{\boldsymbol{u}}^{\mathrm{rx}}$ as the spectrum effectiveness(SE) of UE, $\eta ^{\mathrm{thr}}$ is the minimum SE required to meet Quality of Service (QoS). We simultaneously focus on both the overall performance and fair performance of the codeword region. Thus, the bi-objective optimization problem for generating the codeword $\mathbf{C}_{:,j,k}^{\mathrm{RIS}}$ is
\begin{align}
\label{eq10a}
&\mathcal{P} 1:\max_{\theta _1,\cdots ,\theta _n} \,\mathbb{E} _{\mathcal{A}_k}[\eta _{\boldsymbol{u}}^{\mathrm{rx}}]=\small{\frac{\iint_{\mathcal{A}_k}{\log _2\left( 1+\frac{P_{\boldsymbol{u}}^{\mathrm{rx}}}{\sigma ^2} \right) ds_{\boldsymbol{u}}}}{\iint_{\mathcal{A}_k}{ds_{\boldsymbol{u}}}}},\tag{10a}
\\
&\underset{\theta _1,\cdots ,\theta _n}{\min}P_{\mathcal{A}_k}^{\mathrm{out}}=\mathrm{Pr}_{\mathcal{A}_k}\left( \eta _{\boldsymbol{u}}^{\mathrm{rx}}<\eta ^{\mathrm{thr}} \right) ,\forall \eta ^{\mathrm{thr}}\leqslant \mathbb{E} _{\mathcal{A}_k}[\eta _{\boldsymbol{u}}^{\mathrm{rx}}],\tag{10b}
\label{eq10b}
\\
&\,\,s.t.\ \theta _n\in \mathcal{Q} ,1\leqslant n\leqslant N,\tag{10c}
\label{eq10c}
\end{align}
where ~\eqref{eq10a} represents the objective of maximizing the expected value of the SE as the overall performance metric. ~\eqref{eq10b} represents the objective of minimizing the communication outage rate as the fairness metric. When $\eta ^{\mathrm{thr}} > \mathbb{E} _{\mathcal{A}_k}[\eta _{\boldsymbol{u}}^{\mathrm{rx}}]$, the UE outage rate is inevitably high, stable communication becomes practically impossible. Therefore, we prefer to maintain QoS requirements such that $\eta ^{\mathrm{thr}} \leqslant \mathbb{E} _{\mathcal{A}_k}[\eta _{\boldsymbol{u}}^{\mathrm{rx}}]$ for stable communication. ~\eqref{eq10c} is the quantization constraint of the XL-RISs. 

\section{Problem Solving}
\subsection{Subproblem Definition}
We approximate the coverage performance of the codeword region $\mathcal{A}_k$ by considering a partition into infinitesimal subregions and using the centroids of these subregions in the limit. Then, we define the SE weighted expected optimization problem as
\begin{align}
\label{eq11a}
\mathcal{P} 2:&\max_{\theta _1,\cdots ,\theta _n} \,\,\small{\frac{\sum_{\boldsymbol{u}\in \mathcal{A}_k}{w_{\boldsymbol{u}}\log _2\left( 1+\frac{P_{\boldsymbol{u}}^{\mathrm{rx}}}{\sigma ^2} \right)}}{\sum_{\boldsymbol{u}\in \mathcal{A}_k}{w_{\boldsymbol{u}}}}},\tag{11a}
\\
&\,\,s.t.~\eqref{eq10c},\tag{11b}
\end{align}
where $w_{\boldsymbol{u}} > 0$, $\boldsymbol{u} \in \mathcal{A}_k$ are the weight. In optimization problem $\mathcal{P}2$, due to the large scale of decision variables, we use the Block Coordinate Descent (BCD) method to decompose the original optimization problem into a sequence of sub-optimization problems $\left\{ \mathfrak{p} _1,\mathfrak{p} _2,...,\mathfrak{p} _N\right\}$. We solve them sequentially, with the initial solution for each subproblem being the final solution from the previous subproblem. The general term of the subsequence $\mathfrak{p} _n(1\leqslant n\leqslant N)$ is
\begin{align}
\mathfrak{p} _n:&\max_{\theta _n} \small{\frac{\sum_{\boldsymbol{u}\in \mathcal{A}_k}{w_{\boldsymbol{u}}\log _2\left( 1+\frac{P_{\boldsymbol{u}}^{\mathrm{rx}}}{\sigma ^2} \right)}}{\sum_{\boldsymbol{u}\in \mathcal{A}_k}{w_{\boldsymbol{u}}}}},\tag{12a}
\label{eq12a}
\\
&\,\,s.t.\ \theta _n \in \mathcal{Q}. \tag{12b}
\end{align}

\subsection{Subproblem Solution (closed-form)}
To facilitate further simplification of $P_{\boldsymbol{u}}^{\mathrm{rx}}$, we first define the intermediate variables $\varUpsilon _{\left( n,\boldsymbol{u} \right)}$, $\upsilon _{\left( n,\boldsymbol{u} \right)}$, $\varPsi _{\left( n,\boldsymbol{u} \right)}$, and $\psi _{\left( n,\boldsymbol{u} \right)}$ as
\begin{equation}
\label{eq13}
\left\{ \begin{array}{l}
	\varUpsilon _{\left( n,\boldsymbol{u} \right)}\overset{def}{=}\left| \sum_{i=1,i\ne n}^N{\left[ \mathbf{GC}_{:,j}^{\mathrm{BS}}\mathbf{h\Phi } \right] _{i,i}} \right|,\\
	\upsilon _{\left( n,\boldsymbol{u} \right)}\overset{def}{=}\angle \left( \sum_{i=1,i\ne n}^N{\left[ \mathbf{GC}_{:,j}^{\mathrm{BS}}\mathbf{h\Phi } \right] _{i,i}} \right) ,\\
	\varPsi _{\left( n,\boldsymbol{u} \right)}\overset{def}{=}\left| \left[ \mathbf{GC}_{:,j}^{\mathrm{BS}}\mathbf{h\Phi } \right] _{n,n} \right|,\\
	\psi _{\left( n,\boldsymbol{u} \right)}\overset{def}{=}\angle \left( \left[ \mathbf{GC}_{:,j}^{\mathrm{BS}}\mathbf{h} \right] _{n,n} \right) .\\
\end{array} \right. \tag{13}
\end{equation}
In ~\eqref{eq6}, considering that the term within the modulus operation is a complex value, we can transform it into
\begin{align}
\label{eq14}
P_{\boldsymbol{u}}^{\mathrm{rx}}=P^{\mathrm{tx}}\left| \mathrm{tr}\left( \mathbf{G}\mathbf{C}_{:,j}^{\mathrm{BS}}\mathbf{h}\mathbf{\Phi } \right) \right|^2.\tag{14}
\end{align}
Then, we split it according to the diagonal elements in $\mathbf{\Phi}$, and simplify it by combining \eqref{eq13} to get
\begin{align}
\label{eq15}
P_{\boldsymbol{u}}^{\mathrm{rx}}=P^{\mathrm{tx}}\left| \varUpsilon_{\left( n,\boldsymbol{u} \right)}e^{j\upsilon  _{\left( n,\boldsymbol{u} \right)}}+\varPsi _{\left( n,\boldsymbol{u} \right)}e^{j\left( \psi _{\left( n,\boldsymbol{u} \right)}+\theta _n \right)} \right|^2.
\tag{15}
\end{align}
Finally, using the cosine theorem for complex domains, $P_{\boldsymbol{u}}^{\mathrm{rx}}$ is deconstructed as
\begin{align*}
\label{eq16}
\begin{array}{c}
P_{\boldsymbol{u}}^{\mathrm{rx}}=P^{\mathrm{tx}}\left[ \varPsi _{\left( n,\boldsymbol{u} \right)}^{2}+\varUpsilon _{\left( n,\boldsymbol{u} \right)}^{2}+2\varPsi _{\left( n,\boldsymbol{u} \right)} \right. \varUpsilon _{\left( n,\boldsymbol{u} \right)}
\\
\left. \cos \left( \theta _n+\psi _{\left( n,\boldsymbol{u} \right)}-\upsilon _{\left( n,\boldsymbol{u} \right)} \right) \right] .
\end{array}\tag{16}
\end{align*}

$\varUpsilon_{\left(n,\boldsymbol{u} \right)}$ is the sum of the power transmitted to point $\boldsymbol{u}$ by all EM units except $u_{n}$. $\Psi_{(n, \boldsymbol{u})}$ is the power transmitted only via $u_n$ to point $\boldsymbol{u}$. Therefore,
\begin{align*}
\label{eq17}
&\varUpsilon _{\left( n,\boldsymbol{u} \right)}\gg \varPsi _{\left( n,\boldsymbol{u} \right)},\\
\Rightarrow&\small{\frac{\varPsi _{\left( n,\boldsymbol{u} \right)}^{2}+\varUpsilon_{\left( n,\boldsymbol{u} \right)}^{2}}{2\varPsi _{\left( n,\boldsymbol{u} \right)}\varUpsilon_{\left( n,\boldsymbol{u} \right)}}}>\frac{1}{2}\small{\frac{\varUpsilon_{\left( n,\boldsymbol{u} \right)}}{\varPsi _{\left( n,\boldsymbol{u} \right)}}}\gg 1,
\\
\Rightarrow &\varPsi _{\left( n,\boldsymbol{u} \right)}^{2}+\varUpsilon_{\left( n,\boldsymbol{u} \right)}^{2}\gg 2\varPsi _{\left( n,\boldsymbol{u} \right)}\varUpsilon_{\left( n,\boldsymbol{u} \right)}.
 \tag{17}
\end{align*}

\begin{algorithm}[!t]
\begin{small}
    \caption{AWBCD Algorithm}
    \KwIn{codeword region $\mathcal{A}_k$,\ BS codeword $\mathbf{C}_{:,j}^{\mathrm{BS}}$ ,\ maximum number of iterations $E^{\max} $,\ $\mathbf{\Phi }^{\mathrm{initial}}$,\ $w^{\mathrm{initial}}\in \mathbb{R} _+$.}
    \KwOut{XL-RISs codeword $\mathbf{C}_{:,j,k}^{\mathrm{RIS}}$.}
    initialization:\ $\mathbf{\Phi }\gets \mathbf{\Phi }^{\mathrm{initial}}$,\ $w_{\boldsymbol{u}}\gets w^{\mathrm{initial}}\left( \boldsymbol{u}\in \mathcal{A} _k \right)$.\\
    \For{$e=1\sim E^{\max} $}
    {
        \For{$n=1\sim N$}
        {
            Calculate $\varUpsilon _{\left( n,\boldsymbol{u} \right)})$, $\upsilon _{\left( n,\boldsymbol{u} \right)}$, $\varPsi _{\left( n,\boldsymbol{u} \right)}$, $\psi _{\left( n,\boldsymbol{u} \right)}$ by ~\eqref{eq13}.\\
            Judge $\varsigma$ by ~\eqref{eq18b}.\\
            Calculate $D$ and $U$ by ~\eqref{eq18d} and ~\eqref{eq18e}.\\
            Calculate $\chi$ by ~\eqref{eq18c}.\\
            Update $\theta _n$ by ~\eqref{eq19}.\\
        }
        Update $w_{\boldsymbol{u}}$, $\boldsymbol{u} \in \mathcal{A}_k$ by ~\eqref{eq20}.
    }
    $\mathbf{C}_{:,j,k}^{\mathrm{RIS}}\gets \mathrm{diag}  \left(\mathbf{\Phi } \right).$
\end{small}
\end{algorithm}

Combing ~\eqref{eq16} and \eqref{eq17}, the objective function~\eqref{eq12a} is equivalent to finding the maximum value of the trigonometric function~\eqref{eq18a}.
\begin{align*}
\max_{\theta _n}&\ \  \varsigma \times \sqrt{D^2+U^2}\sin \left( \theta _n+\chi \right),\tag{18a}
\label{eq18a}
\\
&\varsigma =\begin{cases}
	+1,if\,\,\sum_{\boldsymbol{u}\in \mathcal{A}_k}{D\leqslant 0},\\
	-1,if\,\,\sum_{\boldsymbol{u}\in \mathcal{A}_k}{D}>0,\\
\end{cases}\tag{18b}
\label{eq18b}
\\
&\chi =\mathrm{arc}\tan \left( -\frac{U}{D} \right),\tag{18c}
\label{eq18c}
\\
D=&\sum_{\boldsymbol{u}\in \mathcal{A}_k}{\left[ \frac{w_{\boldsymbol{u}}\varPsi _{\left( n,\boldsymbol{u} \right)}\varUpsilon_{\left( n,\boldsymbol{u} \right)}}{\varPsi _{\left( n,\boldsymbol{u} \right)}^{2}+\varUpsilon_{\left( n,\boldsymbol{u} \right)}^{2}}\sin \left( \psi _{\left( n,\boldsymbol{u} \right)}-\upsilon  _{\left( n,\boldsymbol{u} \right)} \right) \right]},\tag{18d}
\label{eq18d}
\\
U=&\sum_{\boldsymbol{u}\in \mathcal{A}_k}{\left[ \frac{w_{\boldsymbol{u}}\varPsi _{\left( n,\boldsymbol{u} \right)}\varUpsilon_{\left( n,\boldsymbol{u} \right)}}{\varPsi _{\left( n,\boldsymbol{u} \right)}^{2}+\varUpsilon_{\left( n,\boldsymbol{u} \right)}^{2}}\cos \left( \psi _{\left( n,\boldsymbol{u} \right)}-\upsilon  _{\left( n,\boldsymbol{u} \right)} \right) \right]}.\tag{18e}
\label{eq18e}
\end{align*}
The $\varsigma$, $\chi$, $D$, and $U$ in \eqref{eq18a} are given by \eqref{eq18b}, \eqref{eq18c}, \eqref{eq18d} and \eqref{eq18e}, respectively. The detailed mathematical derivation is given in the Appendix~\ref{Appendix}. Finally, the solution to the suboptimization problem $\mathfrak{p}_n$ is uniquely determined by
\begin{align*}
\label{eq19}
\theta _n=\begin{cases}
	\underset{\theta \in \mathcal{Q}}{\mathrm{arg}\min}\left| \theta -\left( \small{\frac{\pi}{2}}-\chi  \right) \right|,\ \ if\,\,\varsigma =1,\\
	 \underset{\theta \in \mathcal{Q}}{\mathrm{arg}\min}\left| \theta -\left( \small{\frac{3\pi}{2}}-\chi  \right) \right|,\ \ if\,\,\varsigma =-1.\\
\end{cases}\tag{19}
\end{align*}

\addtolength{\topmargin}{\dimexpr 2.57cm - 1in\relax}
\subsection{Solution Algorithm}
To solve $\mathcal{P}1$, we propose the adaptive weight block coordinate descent (AWBCD) algorithm. It consists of two loops: the inner loop is used to solve the weighted expectation optimization problem $\mathcal{P}2$, and the outer loop dynamically updates the weight coefficients $w_{\boldsymbol{u}}$ to assign greater weight to locations with relatively weaker signals. The weight update strategy is designed as
\begin{align}
w_{\boldsymbol{u}}\gets w_{\boldsymbol{u}}+1,\ \boldsymbol{u} \in \mathcal{A}_k,\ if\,\,\eta _{\boldsymbol{u}}^{\mathrm{rx}}\leqslant \mathbb{E} _{\mathcal{A}_k}[\eta _{\boldsymbol{u}}^{\mathrm{rx}}].
 \tag{20}
\label{eq20}
\end{align}

We have given the closed-form solution for the sub-optimization problem $\mathfrak{p}_n$ in~\eqref{eq19}. In fact, our codebook is designed off-line in advance by the AWBCD algorithm, so the time complexity of the algorithm does not affect the real-time performance of the beam coverage. Furthermore, we designed the BCD algorithm, which does not continuously update the weights as in the AWBCD algorithm. In the subsequent simulations, we will use it as a benchmark to measure the improvement on the fairness performance. Since we place no geometric constraints on $\mathcal{A}_k$, the AWBCD and BCD algorithms are theoretically capable of covering any codeword region within a two-dimensional area using variable-width beams. 

\section{Simulation Comparative Analysis}
We assume that the room in Fig. \ref{Fig2} has dimensions of $20m\times 20m\times 5m$ along the X, Y, and Z axes. Three transmissive XL-RISs are positioned on the wall, with center coordinates $\mathrm{ris}_1(0, 5, 3.5) \mathrm{m}$, $\mathrm{ris}_2(0, 10, 3.5) \mathrm{m}$, and $\mathrm{ris}_3(0, 15, 3.5) \mathrm{m}$. All XL-RISs are uniform planar arrays composed of $200\times200$ EM units. The size of each EM unit is $d_y=d_z=\lambda /2$, and the radiation gain is $G^{\mathrm{RIS}}=8\approx 9.03\mathrm{dBi}$. The normalized power radiation pattern is given by~\cite{9206044}
\begin{align*}
\label{eq21}
&F_{n,m}^{\mathrm{r}}=\begin{cases}
	\cos ^3\theta _{n,m}^{\mathrm{r}},\theta _{n,m}^{\mathrm{r}}\in \left[ 0,\frac{\pi}{2} \right) ,\varphi _{n,m}^{\mathrm{r}}\in \left[ 0,2\pi \right)\\
	0,\ \ \ \ \ \ \ \ \ \theta _{n,m}^{\mathrm{r}}\in \left[ \frac{\pi}{2},\pi \right) ,\varphi _{n,m}^{\mathrm{r}}\in \left[ 0,2\pi \right),\\
\end{cases}\tag{21}\\
&F_{n,\boldsymbol{u}}^{\mathrm{t}}=\begin{cases}
	\cos ^3\theta _{n,\boldsymbol{u}}^{\mathrm{t}},\theta _{n,\boldsymbol{u}}^{\mathrm{t}}\in \left[ 0,\frac{\pi}{2} \right) ,\varphi _{n,\boldsymbol{u}}^{\mathrm{t}}\in \left[ 0,2\pi \right),\\
	0,\ \ \ \ \ \ \ \ \ \textbf{}\theta _{n,\boldsymbol{u}}^{\mathrm{t}}\in \left[ \frac{\pi}{2},\pi \right) ,\varphi _{n,\boldsymbol{u}}^{\mathrm{t}}\in \left[ 0,2\pi \right),
\end{cases}\tag{22}
\end{align*}
where $\theta_{n,m}^{\mathrm{r}}$ and $\varphi_{n,m}^{\mathrm{r}}$ denote the elevation and azimuth angles of the $m$-th BS antenna relative to $u_n$, while $\theta_{n,\boldsymbol{u}}^{\mathrm{t}}$ and $\varphi_{n,\boldsymbol{u}}^{\mathrm{t}}$ represent the corresponding elevation and azimuth angles of $\boldsymbol{u}$ relative to $u_n$. We have labeled these angles in Fig. \ref{Fig3}. Since the beamforming strategy on the side of the BS is not closely related to the research content, we consider a simplified configuration for the BS. Suppose that BS has 3 antennas with coordinates $(-5-\lambda /2,-5,10)m$, $(-5,-5,10)m$, and $(-5+\lambda /2,-5,10)m$. The total transmit power $P^{\mathrm{tx}}=44\mathrm{dbm}$. BS and UE are equipped with isotropic antennas, $G^{\mathrm{tx}}=G^{r\mathrm{x}}=1=0\mathrm{dBi}$, $F_{n,m}^{t\mathrm{x}}=F_{n,\boldsymbol{u}}^{\mathrm{rx}}=1=0\mathrm{dBi}$. The Gaussian white noise power is $\sigma ^2=-105\mathrm{dbm}$. The frequency of the electromagnetic wave is $30\mathrm{GHz}$. We assume that $\mathbf{C}_{:,1}^{\mathrm{BS}}=\frac{1}{\sqrt{3}}\left[ 1,1,1 \right] ^{\mathrm{T}}$ is a codeword from the commonly used DFT codebook~\cite{3GPP.TS.38.214.2018}. The initial coefficient matrix $\mathbf{\Phi}^{\mathrm{initial}}$ of the XL-RIS is randomly generated, and the initial weight $w^{\mathrm{initial}}$ is set to 100.

\begin{figure}[!t]
\centering
\setlength{\abovecaptionskip}{-0.32cm}
\includegraphics[width=2.8in]{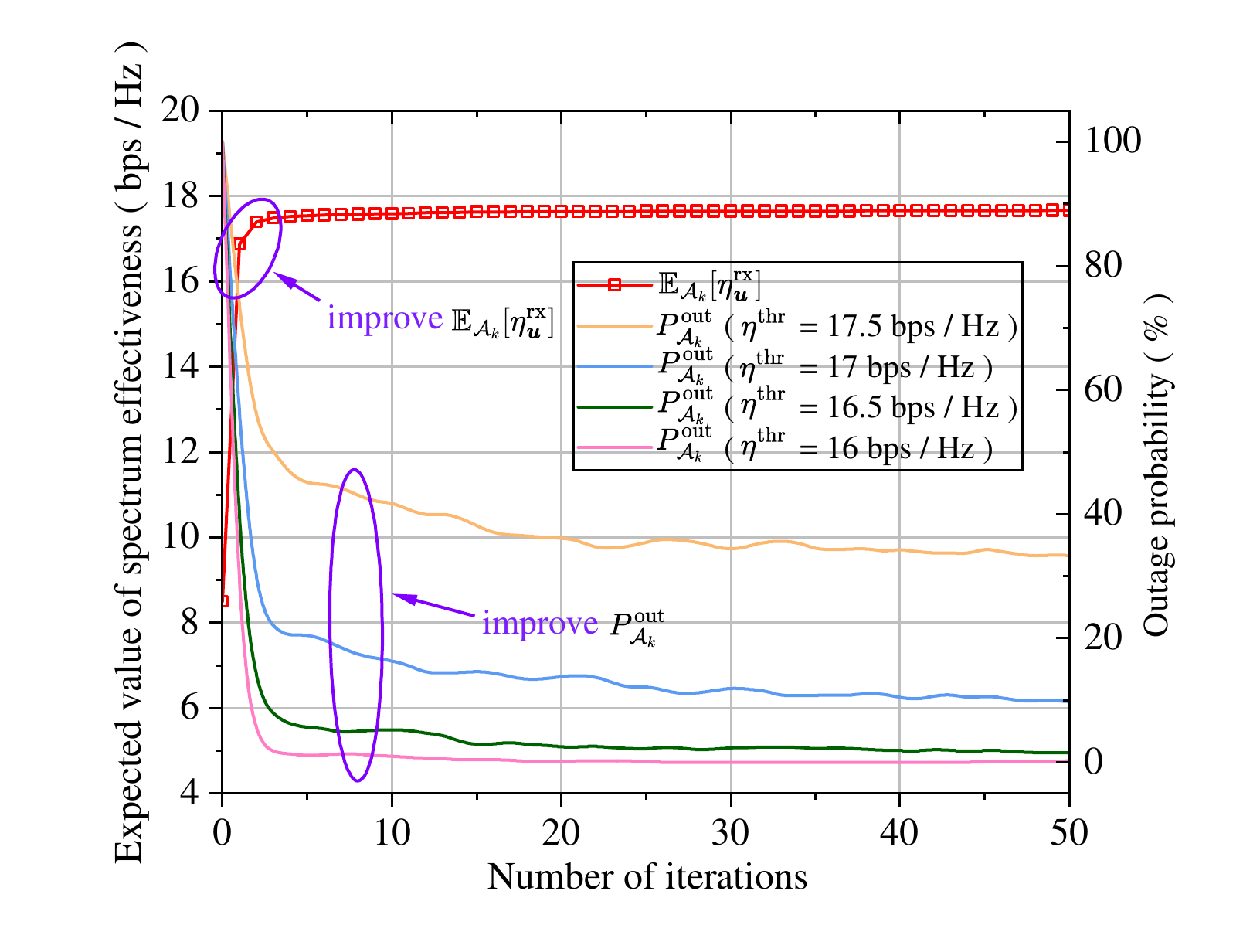}
\caption{Iteration plot for solving the codeword $\mathbf{C}_{:,1,k}^{\mathrm{RIS}}$ using the AWBCD algorithm. $\mathcal{A}_k=\left\{ \left( x,y,0.5 \right) m|x,y\in \left( 8.5,11.5 \right) \right\}$.}
\label{Fig5}
\vspace{-0.8cm}
\end{figure}

\begin{figure} [t!]
	\centering
    \captionsetup[subfigure]{justification=centering}
	\subfloat[$x,y\in \left( 8.5,11.5 \right) $]{
 		\includegraphics[scale=0.195]{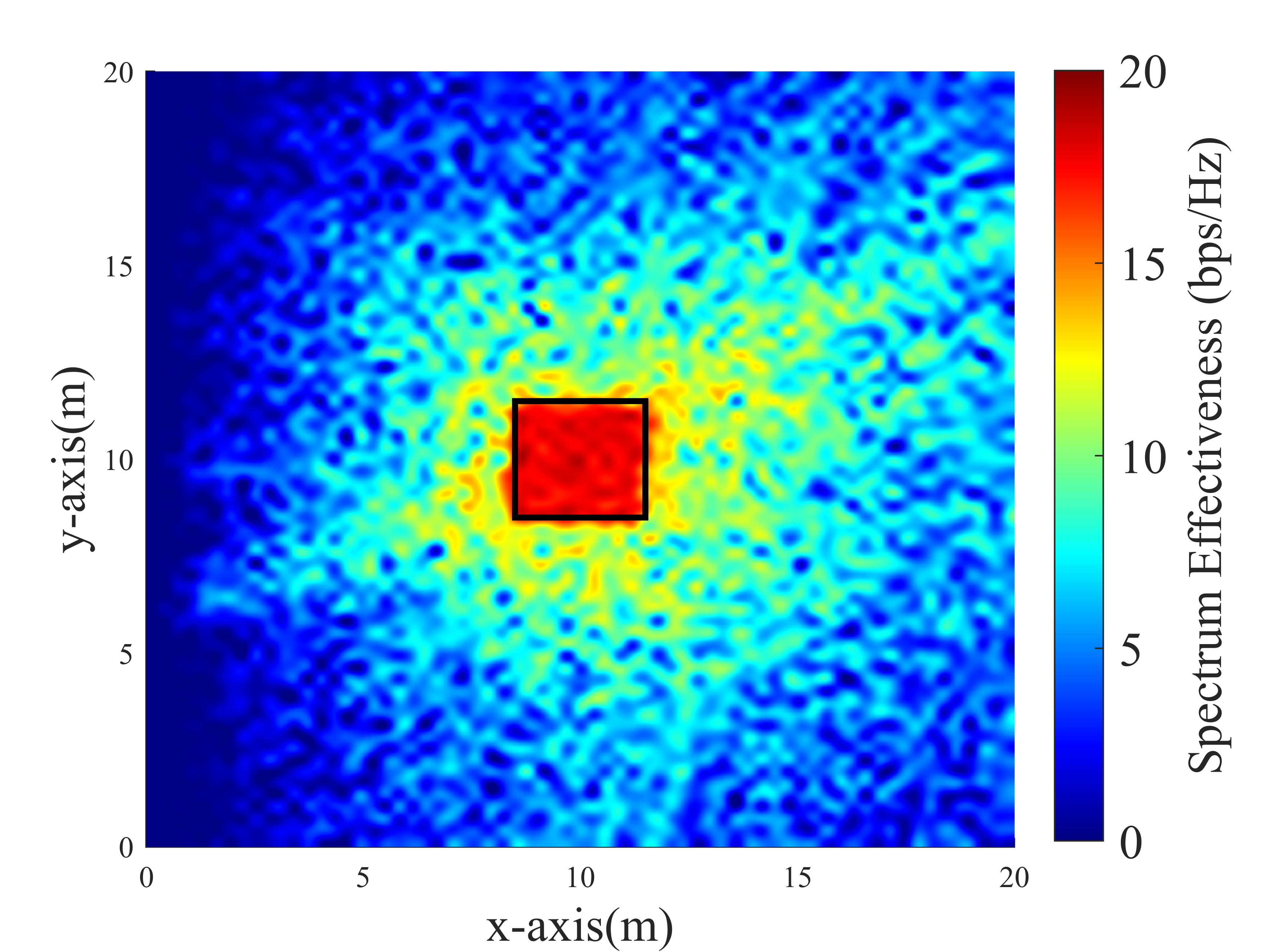}\label{subfig:a}}
	\subfloat[$x,y\in \left( 7.5,12.5 \right) $]{
		\includegraphics[scale=0.195]{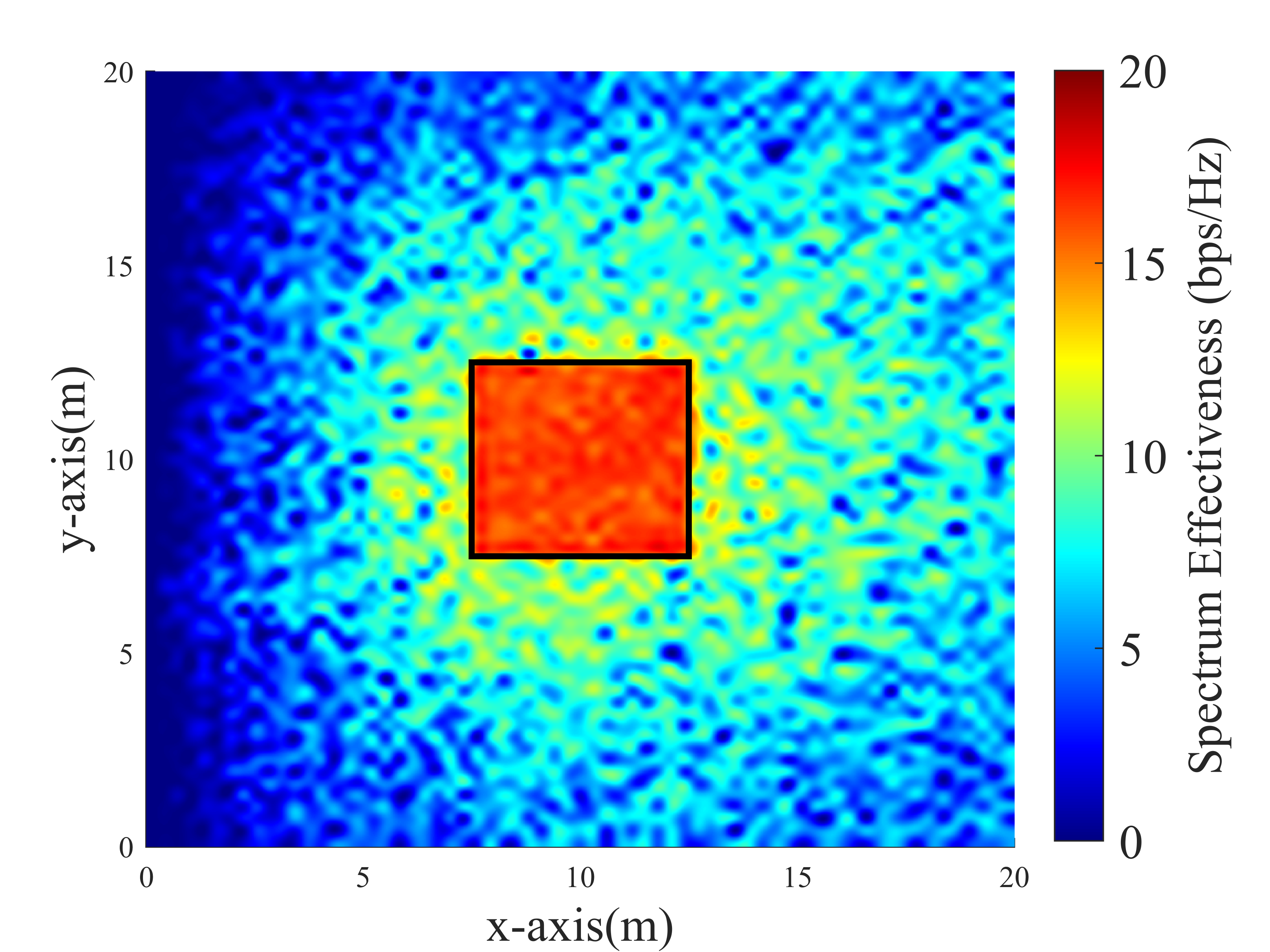}\label{subfig:b}}
	\subfloat[$x,y\in \left( 3.5,6.5 \right) $]{
		\includegraphics[scale=0.195]{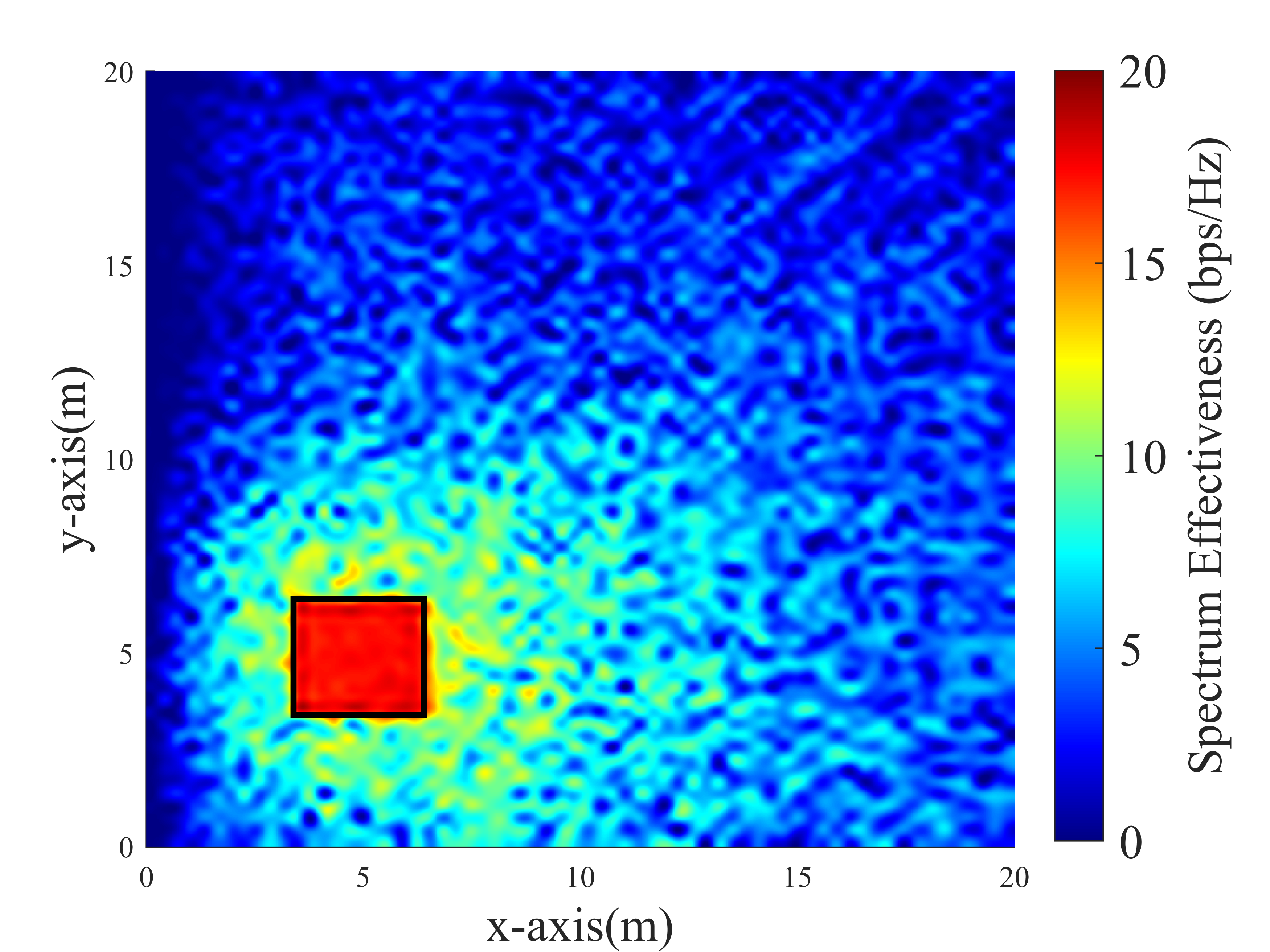}\label{subfig:c}}
        \\
        \vspace{-10pt}
    \subfloat[$x,y\in \left( 3.5,6.5 \right) \cup \left( 13.5,16.5 \right)$]{
	  \includegraphics[scale=0.195]{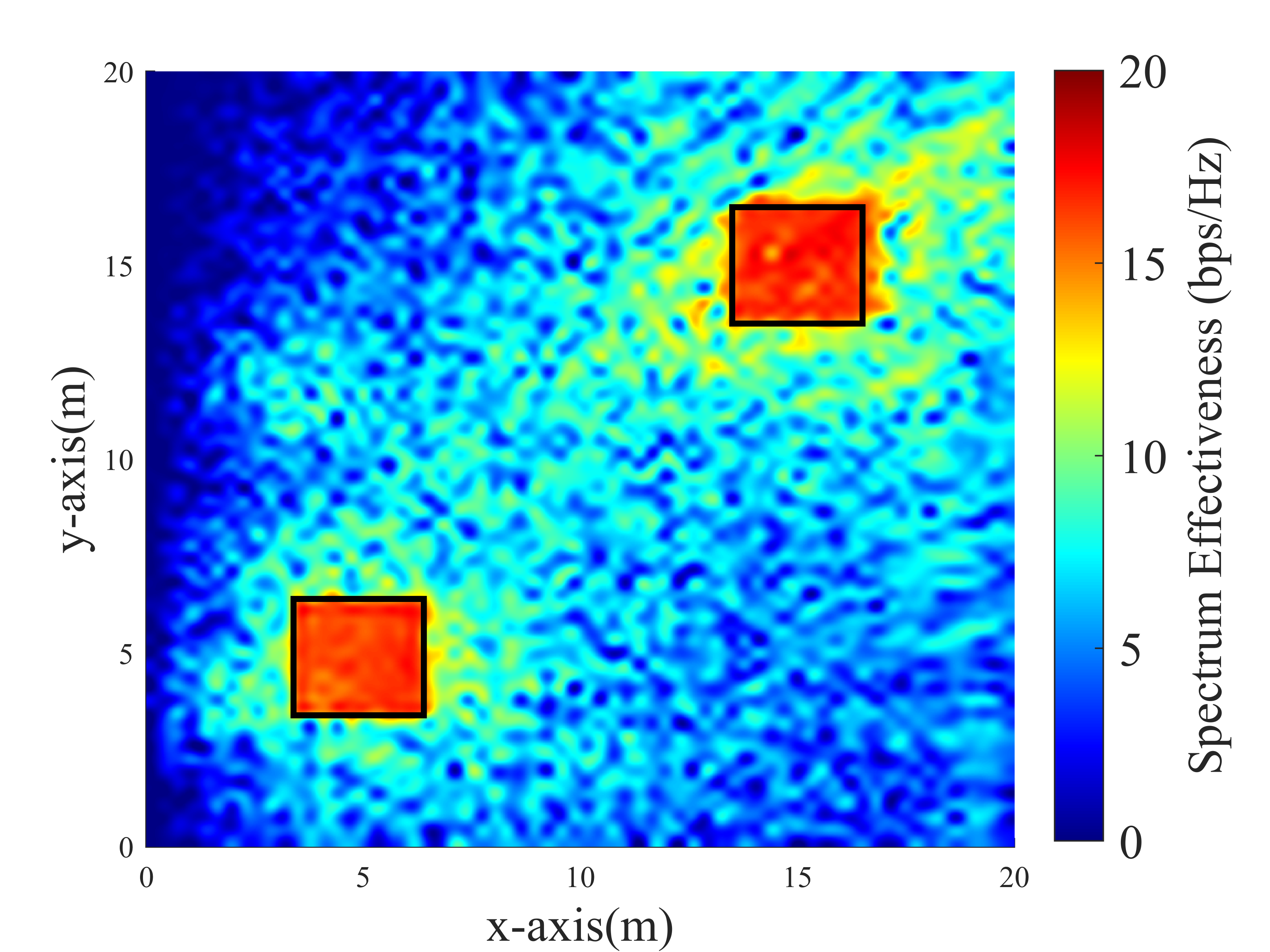}\label{subfig:d}}
      \subfloat['$\mathrm{T}$'-shaped codeword region]{
	  \includegraphics[scale=0.195]{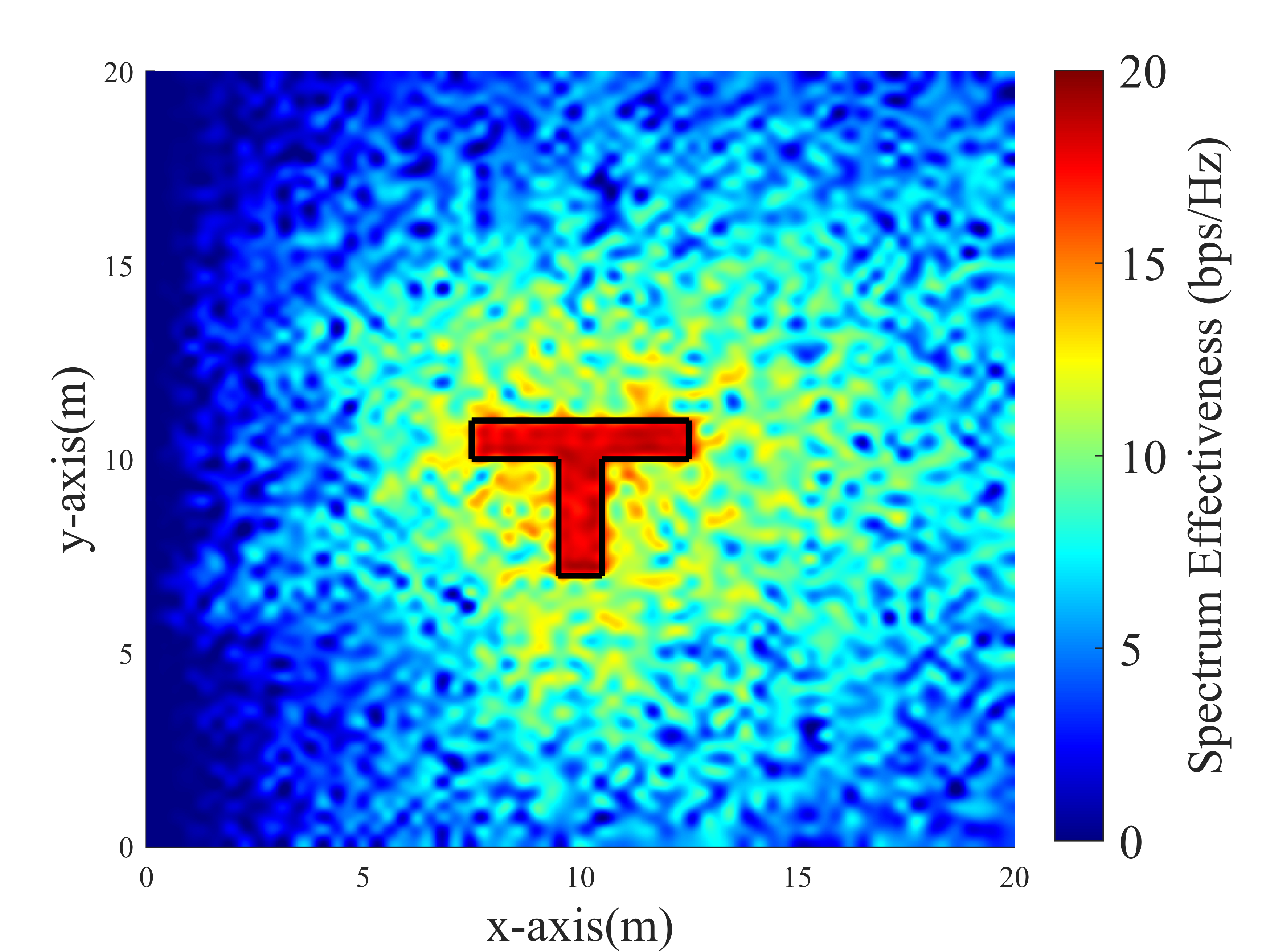}\label{subfig:e}}
      \subfloat['$\mathrm{L}$'-shaped codeword region]{
	  \includegraphics[scale=0.195]{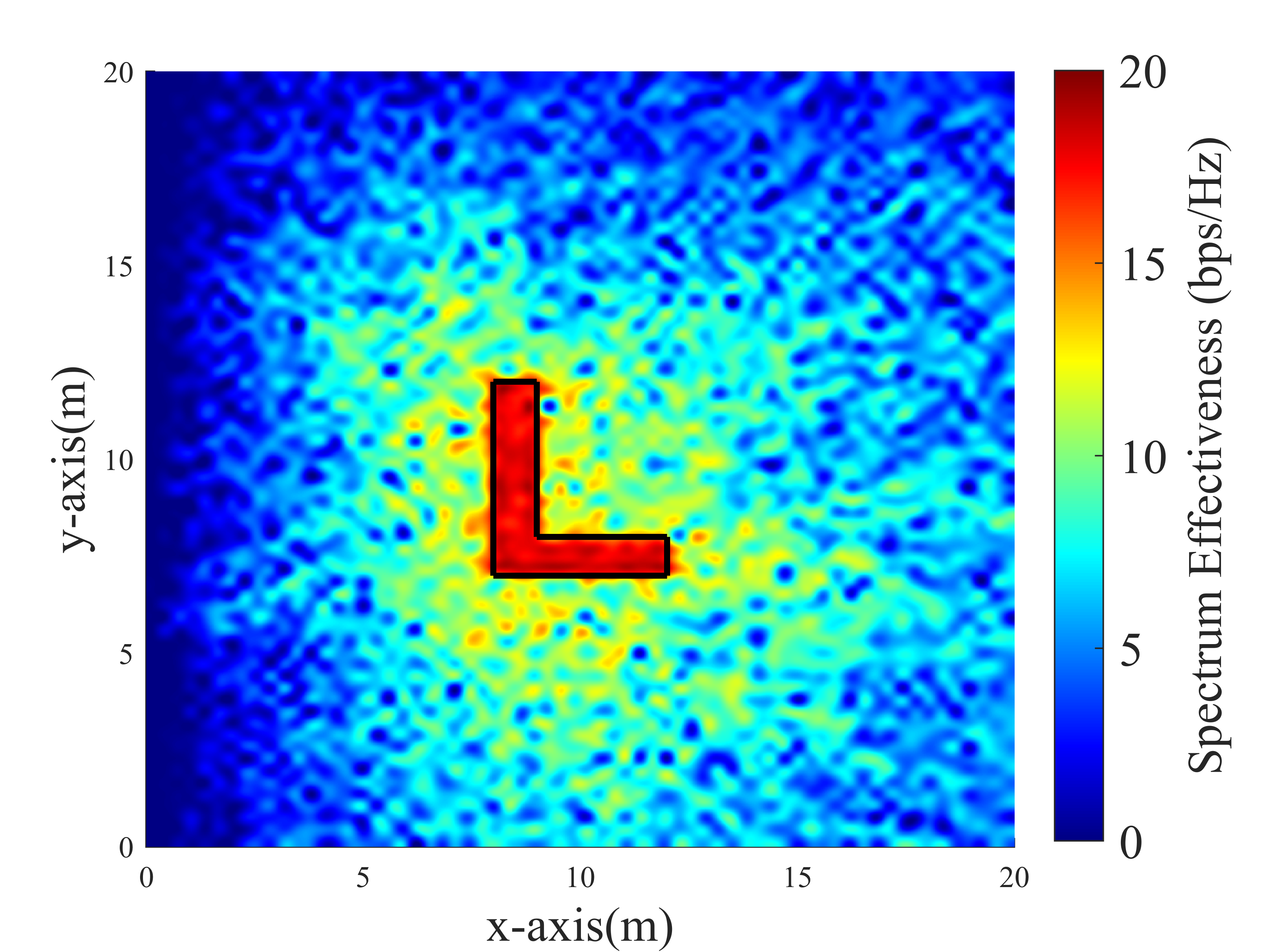}\label{subfig:f}}
    \caption{SE heatmaps for different codeword region coverages. $\mathcal{A}_k=\left\{ \left( x,y,0.5 \right) m \right\}$.}
\label{Fig6}
\vspace{-0.05cm}
\end{figure}

Fig. \ref{Fig5} shows the iteration diagram for designing the codeword $\mathbf{C}_{:,1,k}^{\mathrm{RIS}}$ using the AWBCD algorithm. We observe that $\mathbb{E} _{\mathcal{A}_k}[\eta _{\boldsymbol{u}}^{\mathrm{rx}}]$ within the codeword region monotonically increases and rapidly converges with algorithm iterations, and the $P_{\mathcal{A}_k}^{\mathrm{out}}$ for different $\eta ^{\mathrm{thr}}$ generally decreases and converges with algorithm iterations. In the early iterations, the algorithm focuses more on the overall performance metric $\mathbb{E} _{\mathcal{A}_k}[\eta _{\boldsymbol{u}}^{\mathrm{rx}}]$. In the mid-to-late iterations, the algorithm focuses on fairness metrics, optimizing sub-regions with relatively weak signals while ensuring overall performance is not reduced, aiming to reduce $P_{\mathcal{A}_k}^{\mathrm{out}}$.

\begin{figure}[!t]
\centering
\setlength{\abovecaptionskip}{-0.32cm}
\includegraphics[width=2.65in]{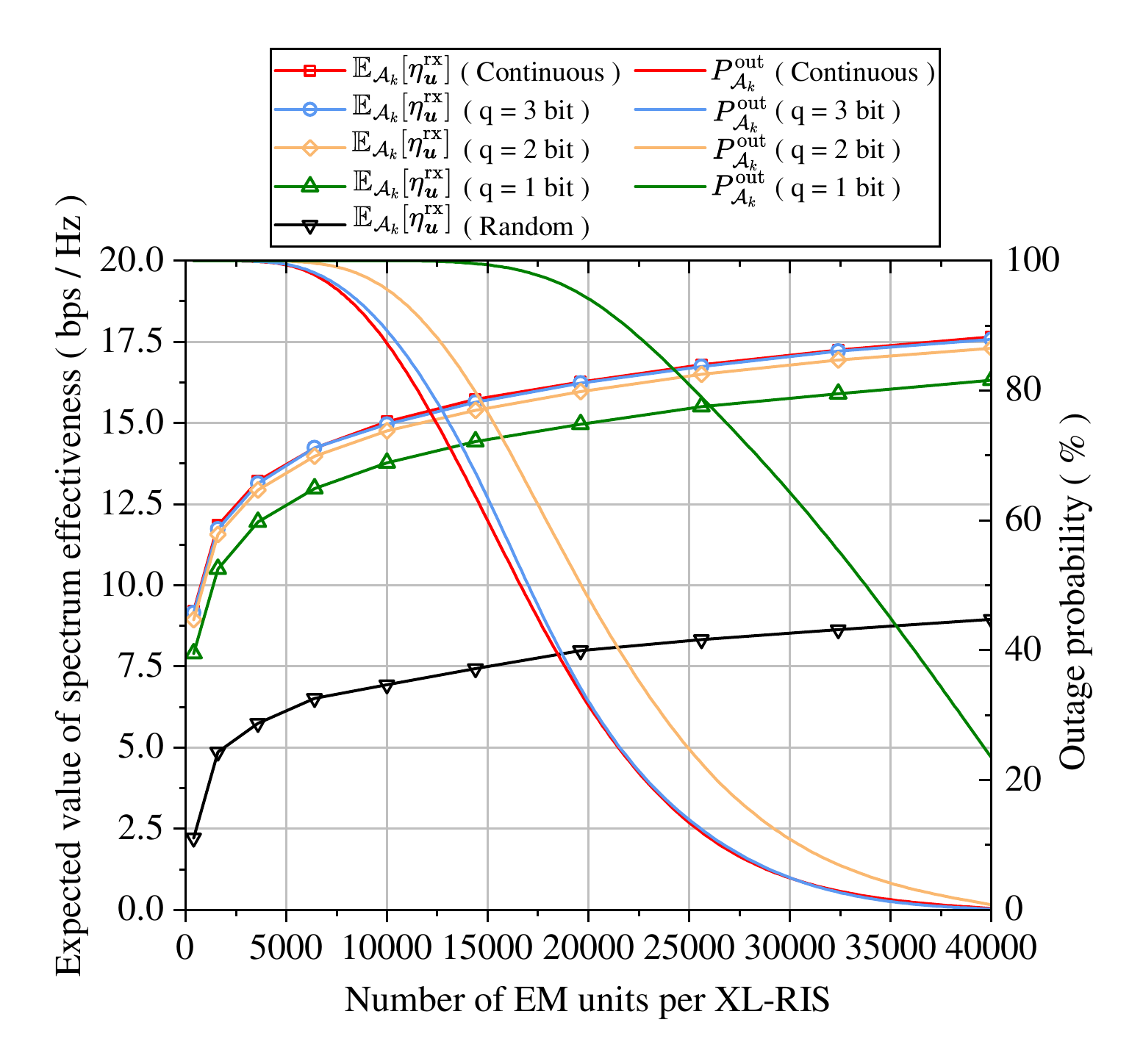}
\caption{Coverage performance comparison for different numbers of EM units. $\eta ^{\mathrm{thr}}$=16bps/Hz. $\mathcal{A}_k=\left\{ \left( x,y,0.5 \right) m|x,y\in \left( 8.5,11.5 \right) \right\}$. }
\label{Fig7}
\vspace{-0.77cm}
\end{figure}

\begin{figure}[t!]
	\centering
    \captionsetup[subfigure]{justification=centering, skip=-8pt}  
	\subfloat[$r\in \left( 8,10 \right) ,\xi \in \left( -5\degree,5\degree \right) $.]{
		\includegraphics[scale=0.185]{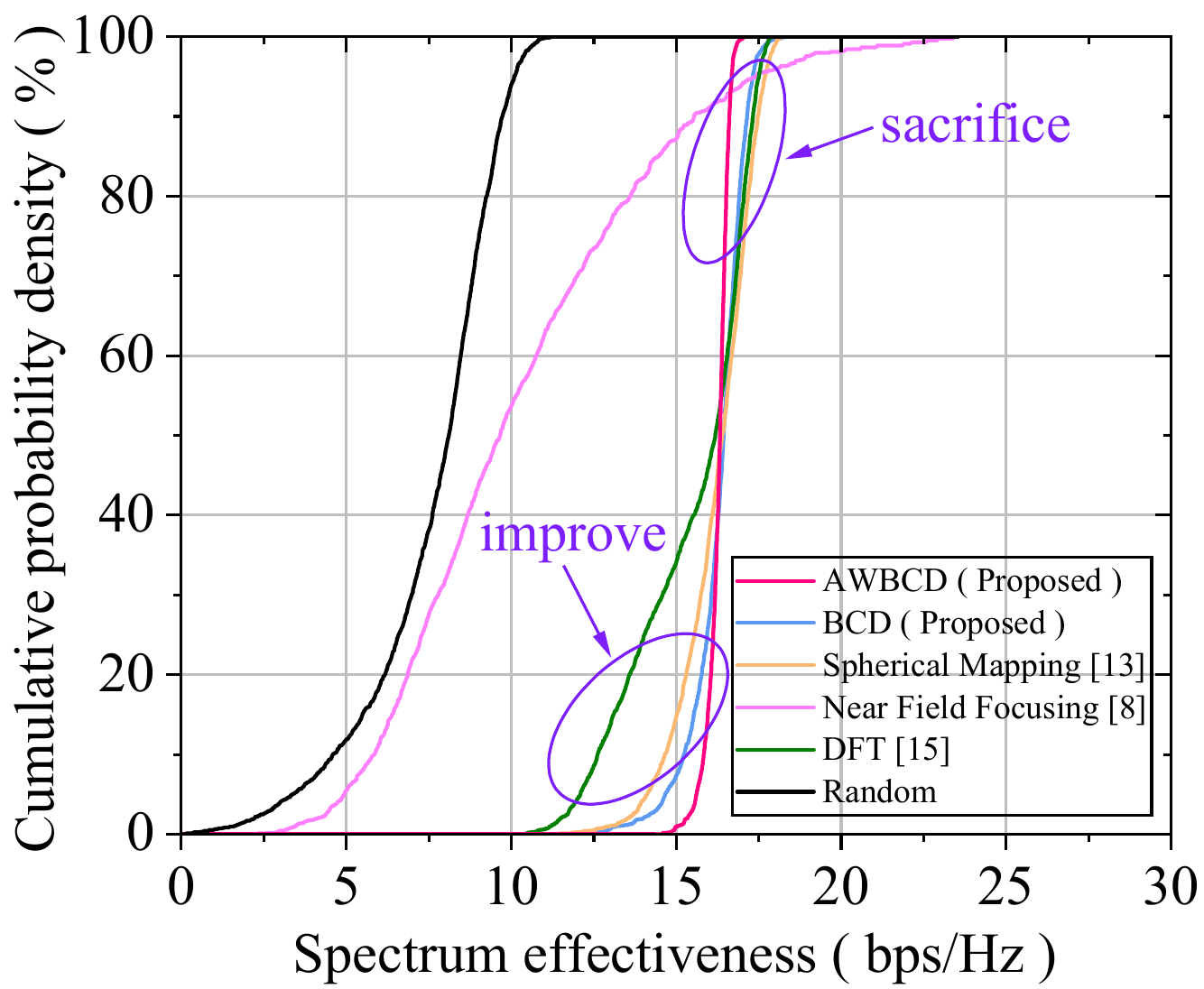}
		\vspace{-0.2cm}
		\label{subfig:a}
	}
	\subfloat[$r\in \left( 8,10 \right) ,\xi \in \left( -65\degree,-55\degree \right) $.]{
		\includegraphics[scale=0.185]{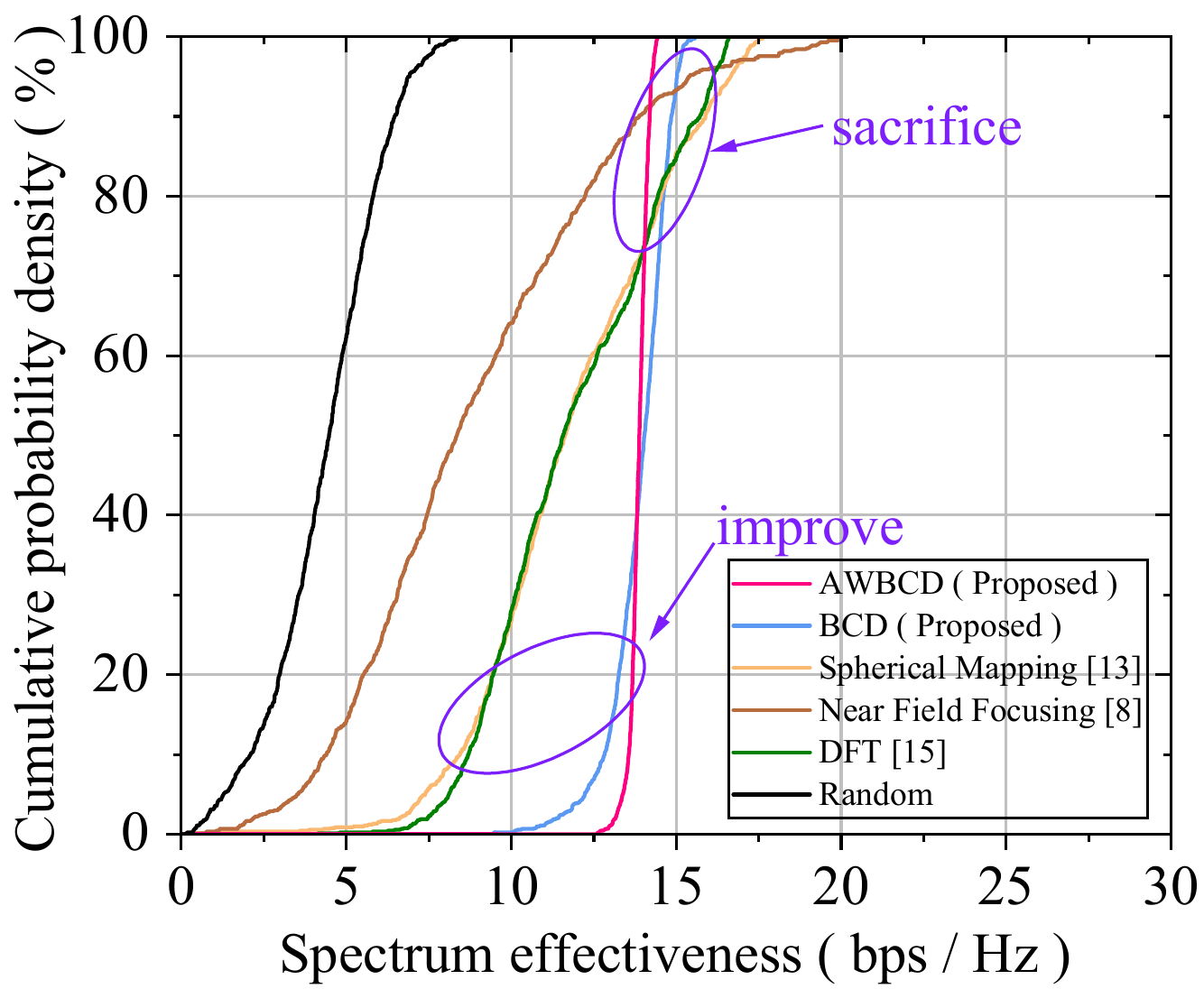}
		\vspace{-0.2cm}
		\label{subfig:b}
	}
	\caption{Plot of the cumulative probability density distribution of the spectral efficiency of the codeword region. $\mathcal{A}_k =\left\{ \left( r\cos \xi ,r\sin \xi ,0.5 \right) m \right\} $.}
	\label{Fig8}
	\vspace{-0.1cm}
\end{figure}

Fig. \ref{Fig6}(a) is the SE heatmap for the corresponding codeword region. UEs at any position within the codeword region ${\mathcal{A}_k}$ have relatively high SE. Fig. \ref{Fig6}(b-d) sequentially show the coverage performance when the codeword region is widened, shifted, and in multi-target scenarios. We observe that the codewords designed by the AWBCD algorithm achieve good coverage in different codeword regions, demonstrating the robustness of the algorithm. Fig. {\ref{Fig6}(e-f)} respectively demonstrate SE heatmaps of beam coverage for 'T'-shaped and 'L'-shaped codeword regions using the AWBCD algorithm, which validates the applicability of the algorithm to unconventional codeword region shapes.

Fig. \ref{Fig7} shows the variation of $\mathbb{E} _{\mathcal{A}_k}[\eta _{\boldsymbol{u}}^{\mathrm{rx}}]$ and $P_{\mathcal{A}_k}^{\mathrm{out}}$ with the number of EM units. We observe that increasing the number of EM units can significantly improve the signal coverage performance of the codeword region and reduce the outage rate. Furthermore, the phase quantization bits of the XL-RIS also affect the coverage performance. When $q=1$ bit, there is a noticeable degradation in coverage performance. When $q \geqslant 3$ bits, its performance is essentially consistent with that of XL-RIS with a continuous phase shift.

\begin{figure*}[t!]
\begin{scriptsize}
\centering
\begin{align*}
    \label{eqA1}
        ~\eqref{eq12a}&\approx  \max_{\theta _n} \prod_{\boldsymbol{u}\in \mathcal{A}_k}{\left( P_{\boldsymbol{u}}^{\mathrm{rx}} \right) ^{w_{\boldsymbol{u}}}} \tag{A1} \\
    \label{eqA2}
	&=  \underset{\theta _n}{\max}\left( P^{\mathrm{tx}} \right) ^{\sum_{\boldsymbol{u}\in \mathcal{A}_k}{w_{\boldsymbol{u}}}}\prod_{\boldsymbol{u}\in \mathcal{A}_k}{\left[ \varPsi _{\left( n,\boldsymbol{u} \right)}^{2}+\varUpsilon _{\left( n,\boldsymbol{u} \right)}^{2}+2\varPsi _{\left( n,\boldsymbol{u} \right)} \right.}\left. \varUpsilon _{\left( n,\boldsymbol{u} \right)}\cos \left( \theta _n+\psi _{\left( n,\boldsymbol{u} \right)}-\upsilon _{\left( n,\boldsymbol{u} \right)} \right) \right] ^{w_{\boldsymbol{u}}} \tag{A2} \\
    \label{eqA3}
	&= \max_{\theta _n} \prod_{\boldsymbol{u}\in \mathcal{A}_k}{\sum_{r=0}^{\infty}{\left\{ C_{w_{\boldsymbol{u}}}^{r}\left[ \varPsi _{\left( n,\boldsymbol{u} \right)}^{2}+\varUpsilon _{\left( n,\boldsymbol{u} \right)}^{2} \right] ^{w_{\boldsymbol{u}}-r}\left[ 2\varPsi _{\left( n,\boldsymbol{u} \right)}\varUpsilon _{\left( n,\boldsymbol{u} \right)}\cos \left( \theta _n+\psi _{\left( n,\boldsymbol{u} \right)}-\upsilon _{\left( n,\boldsymbol{u} \right)} \right) \right] ^r \right\}}} \tag{A3} \\
    \label{eqA4}
	&\approx  \max_{\theta _n} \prod_{\boldsymbol{u}\in \mathcal{A}_k}{\left( \varPsi _{\left( n,\boldsymbol{u} \right)}^{2}+\varUpsilon_{\left( n,\boldsymbol{u} \right)}^{2} \right) ^{w_{\boldsymbol{u}}-1}}\prod_{\boldsymbol{u}\in \mathcal{A}_k}{\left[ \varPsi _{\left( n,\boldsymbol{u} \right)}^{2}+\varUpsilon_{\left( n,\boldsymbol{u} \right)}^{2}+2w_{\boldsymbol{u}}\varPsi _{\left( n,\boldsymbol{u} \right)}\varUpsilon_{\left( n,\boldsymbol{u} \right)}\cos \left( \theta _n+\psi _{\left( n,\boldsymbol{u} \right)}-\upsilon  _{\left( n,\boldsymbol{u} \right)} \right) \right]} \tag{A4}\\
    \label{eqA5}
	&=  \max_{\theta _n} \prod_{\boldsymbol{u}\in \mathcal{A}_k}{\left[ \varPsi _{\left( n,\boldsymbol{u} \right)}^{2}+\varUpsilon_{\left( n,\boldsymbol{u} \right)}^{2}+2w_{\boldsymbol{u}}\varPsi _{\left( n,\boldsymbol{u} \right)}\varUpsilon_{\left( n,\boldsymbol{u} \right)}\cos \left( \theta _n+\psi _{\left( n,\boldsymbol{u} \right)}-\upsilon  _{\left( n,\boldsymbol{u} \right)} \right) \right]} \tag{A5}\\
    \label{eqA6}
	&\approx \max_{\theta _n} \sum_{\boldsymbol{u}\in \mathcal{A}_k}{\left\{ \frac{w_{\boldsymbol{u}}\varPsi _{\left( n,\boldsymbol{u} \right)}\varUpsilon_{\left( n,\boldsymbol{u} \right)}}{\varPsi _{\left( n,\boldsymbol{u} \right)}^{2}+\varUpsilon_{\left( n,\boldsymbol{u} \right)}^{2}}\cos \left( \theta _n+\psi _{\left( n,\boldsymbol{u} \right)}-\upsilon  _{\left( n,\boldsymbol{u} \right)} \right) \right\}}\tag{A6}\\
    \label{eqA7}
	&=\max_{\theta _n} \left\{\sum_{\boldsymbol{u}\in \mathcal{A} _k}{\left[ \frac{w_{\boldsymbol{u}}\varPsi _{\left( n,\boldsymbol{u} \right)}\varUpsilon _{\left( n,\boldsymbol{u} \right)}}{\varPsi _{\left( n,\boldsymbol{u} \right)}^{2}+\varUpsilon _{\left( n,\boldsymbol{u} \right)}^{2}}\cos \left( \psi _{\left( n,\boldsymbol{u} \right)}-\upsilon _{\left( n,\boldsymbol{u} \right)} \right) \right]}\cos \theta _n-\sum_{\boldsymbol{u}\in \mathcal{A} _k}{\left[ \frac{w_{\boldsymbol{u}}\varPsi _{\left( n,\boldsymbol{u} \right)}\varUpsilon _{\left( n,\boldsymbol{u} \right)}}{\varPsi _{\left( n,\boldsymbol{u} \right)}^{2}+\varUpsilon _{\left( n,\boldsymbol{u} \right)}^{2}}\sin \left( \psi _{\left( n,\boldsymbol{u} \right)}-\upsilon _{\left( n,\boldsymbol{u} \right)} \right) \right] \sin \theta _n} \right\} 
 \tag{A7}\\
    \overline{\ \ \ \ \ \ \ } &\overline{\ \ \ \ \ \ \ \ \ \ \ \ \ \ \ \ \ \ \ \ \ \ \ \ \ \ \ \ \ \ \ \ \ \ \ \ \ \ \ \ \ \ \ \ \ \ \ \ \ \ \ \ \ \ \ \ \ \ \ \ \ \ \ \ \ \ \ \ \ \ \ \ \ \ \ \ \ \ \ \ \ \ \ \ \ \ \ \ \ \ \ \ \ \ \ \ \ \ \ \ \ \ \ \ \ \ \ \ \ \ \ \ \ \ \ \ \ \ \ \ \ \ \ \ \ \ \ \ \ \ \ \ \ \ \ \ \ \ \ \ \ \ \ \ \ \ \ \ \ \ \ \ \ \ \ \ \ \ \ \ \ \ \ \ \ \ \ \ \ \ \ \ \ \ \ \ \ \ \ \ \ \ \ \ \ \ \ \ \ \ \ \ \ \ \ \ \ \ \ \ \ \ \ \ }
\end{align*}
\end{scriptsize}
\vspace{-1.25cm}
\end{figure*}

In existing RIS near-field variable-width codebook designs, the shape of the codeword region is typically fixed as a sector or a rectangle, and these designs are exclusively intended for a single RIS. To ensure fairness of comparison, we assume that only $\mathrm{ris}_2$ at $(0, 10, 3.5)m$ exists, and the codeword region is uniformly designed according to a sector $\mathcal{A}_k =\left\{ \left( r\cos \xi ,r\sin \xi ,0.5 \right) m \right\} $. Fig. \ref{Fig8}(a-b) show the cumulative distribution function (CDF) of the SE for the codeword regions. We observe that the proposed AWBCD codeword and BCD codeword achieve significant improvements in terms of overall performance and fairness compared to other schemes. Especially when the angle between the codeword region and the normal vector RIS is excessively large, spherical mapping codeword ~\cite{10437101} cannot achieve uniform signal coverage, while our codeword maintains good performance. We also find that the AWBCD codeword improves coverage at the edges of the codeword region by sacrificing signal strength in the center of the region. This ultimately successfully reduces the probability of communication outage for users within the codeword region, achieving fair signal coverage.

\section{Conclusion}
In this paper, we innovatively proposed a near-field variable-width beam generation algorithm for XL-RIS and applied it to codebook design. Simulation results demonstrated that our scheme outperformed existing methods in both overall performance and fairness performance, and exhibited excellent robustness in codeword region variations. Furthermore, our algorithm supported arbitrary-shaped codeword region coverage and multi-XL-RISs joint codebook design, which was unprecedented in existing work. In the future, we will attempt to validate the algorithm using hardware and explore the potential of variable-width codebooks to enhance beam alignment accuracy and reduce scanning overhead, aiming to promote more efficient RIS-assisted communication~\cite{Cui2025AI6G}.

\section{Appendix}\label{Appendix}
We detail the derivation process from ~\eqref{eq12a} to~\eqref{eq18a} above on this page. The approximation in ~\eqref{eqA1} is because we consider $P_{\boldsymbol{u}}^{\mathrm{rx}}\gg \sigma ^2$ and $\sum_{\boldsymbol{u}\in \mathcal{A} _k}{w_{\boldsymbol{u}}}$ is a constant. In ~\eqref{eqA3}, we use the generalized binomial expansion, where $C_{w_{\boldsymbol{u}}}^{r}=\frac{w_{\boldsymbol{u}}(w_{\boldsymbol{u}}-1)(w_{\boldsymbol{u}}-2)\cdots (w_{\boldsymbol{u}}-r+1)}{r!}$. The approximations in ~\eqref{eqA4} and~\eqref{eqA6} are due to considering the inequality~\eqref{eq17}. Therefore, we expand it and approximate it using the largest and second-largest terms. The remaining steps are standard simplifications and rearrangements.

\bibliographystyle{IEEEtran}
\bibliography{references}

\end{document}